# Efficient electrochemical reduction of CO$_2$ to CO by "soft" functional materials

Yanjie Hu [a,b], Jiaqi Feng [a], Xiangping Zhang [a], Hongshuai Gao [a], Saimeng Jin [c], Lei Liu [a,*], Weifeng Shen [c,d*]

[a] Key Laboratory of Green Process and Engineering, State Key Laboratory of Multiphase Complex System, Beijing Key Laboratory of Ionic Liquids Clean Process, Institute of Process Engineering, Chinese Academy of Sciences, Beijing, 100190, China

[b] School of Materials and Energy, Guangdong University of Technology, Guangzhou 510006, China

[c] School of Chemistry and Chemical Engineering, Chongqing University, Chongqing 400044, China

[d] Chongqing Key Laboratory of Theoretical and Computational Chemistry, Chongqing 400044, China

*Corresponding authors:

liulei@ipe.ac.cn, liulei3039@gmail.com (L. Liu)

shenweifeng@cqu.edu.cn (W. Shen).

**Abstract**：Electrochemical reduction of carbon dioxide ($CO_2$) to CO is a promising strategy. However, achieving high Faradaic efficiency with high current density using ionic liquids (ILs) electrolyte remains a challenge. In this study, the new ionic liquids (ILs) *N*-octyltrimethyl 1,2,4-triazole ammonium ($[N_{1118}][TRIZ]$) shows outstanding performance for electrochemical reduction of $CO_2$ to CO on the commercial Ag electrode, and the current density can be up to 50.8 mAcm$^{-2}$ with a Faradaic efficiency of 90.6%. The current density of CO is much higher than those reported in the ILs electrolyte. In addition, the density functional theory (DFT) calculation further proved that $[N_{1118}][TRIZ]$ interact with $CO_2$ to form $[N_{1118}]^+[TRIZ-CO_2]^-$ complex which played a key role in reducing the activation energy of $CO_2$. According to the molecular orbital theory, the electrons obtained from ILs was filled in the anti-bonding orbit ($\pi^*$) of the $CO_2$, resulting in reducing the C=O bond energy. This work provides a new strategy to design novel ILs for high efficiency electrochemical reduction of $CO_2$ to CO.

# 1. Introduction

Carbon dioxide ($CO_2$), as one of most important greenhouse gases, is considered to be a non-toxic, inexpensive, renewable, and infinite C1-feedstock [1]. Converting $CO_2$ into useful products is considered as an important way towards $CO_2$ utilization [2]. An important conversion process is the reduction of $CO_2$ to CO. The product CO is promising as the main reactant of Fischer-Tropsch process [3]. Compared with traditional thermal reaction, the electrochemical reduction is a very promising strategy to convert $CO_2$ to CO under mild conditions by using hydro energy, solar and renewable wind [4, 5]. However, the reduction process faces a significant challenge mainly owing to large recombination energy in the transition of linear $CO_2$ to curved $CO_2^{•-}$ [6]. For this reason, certain electrolyte allowing for formation of complex with $CO_2$ would lower the reorganization energy for $CO_2$ activation [7].

As a 'designer solvent', room temperature ionic liquids (hereinafter referred to as 'ILs') are organic salts made up of anions and cations, which are liquids with temperature below 100 °C [8]. Recently, ILs have received much attention in the electrochemical reduction of $CO_2$ thanks to their properties of negligible vapor pressure, high $CO_2$ solubility, high electric conductivity and a wide potential window [9]. The ILs as promising cocatalysts have achieved impressive selectivity and yield in the electrocatalytic reduction of $CO_2$ to CO [10, 11]. Zhou et al. [12] presented that polycrystalline Ag cathode in ILs 1-butyl-3-methylimidazolium chloride ([EMIM]Cl) showed excellent synergy in electrochemical $CO_2$ reduction to CO with > 99% selectivity. Rosen et al. [6] reported that [EMIM][$BF_4$] could help to provide a low-energy pathway for the electrochemical reduction of $CO_2$ to CO by the formation of [EMIM-$CO_2$]---[$BF_4$] complex. Rudnev et al. [11] reported the combination of [BMIM][$BF_4$] with porous nanostructured silver foam catalyst produced a wide potential window of ~300 mV, where the Faraday efficiency of CO exceeds 94%. The electrocatalytic reduction of $CO_2$ to CO using [$C_n$mim][$BF_4$] ILs has been widely studied, however, the ILs containing [$BF_4$]$^-$ anions are instable towards hydrolysis in contract with moisture, resulting in the formation of toxic and corrosive

HF [13]. Meanwhile, the applicability of $CO_2$ reduction in imidazolium-based ILs is limited due to the low solubility of $CO_2$ [14]. At present, the above problems have been overcome by the use of ILs with 1,2,4-triazole as anion [7, 14]. The anion is able to chemically bind $CO_2$ to form $[TRIZ-CO_2]^-$ complex resulting in high $CO_2$ solubility. However, the ILs are reported difficult to be synthesized. And to date, the use of ILs with 1,2,4-triazole as anion in electrochemical $CO_2$ reduction to CO has not been found. It is frequently encountered that the electrocatalytic $CO_2$ reduction to CO in ILs with satisfactory current density particularly have the electrodes made of commercial materials [15, 16]. Considering the above problems, herein, a new ILs *N*-octyltrimethyl 1,2,4-triazole ammonium ($[N_{1118}][TRIZ]$) is designed and employed in electrocatalytic reduction of $CO_2$. $[N_{1118}][TRIZ]$ has been shown to have high $CO_2$ absorption capacity and is easy to be synthesized [17]. And the important feature of the $[N_{1118}][TRIZ]$ is the anion interact with $CO_2$ to form $[TRIZ-CO_2]^-$ complex, thus reducing the activation energy of $CO_2$. It is particularly interested to discover whether the electrocatalytic $CO_2$ reduction with $[N_{1118}][TRIZ]$ ILs as cocatalyst can bring about satisfactory efficiency.

Therefore, to achieve high Faradaic efficiency with high current density using ionic liquids (ILs) electrolyte. In this study, a new ionic liquid $[N_{1118}][TRIZ]$ was synthesized by one-step method and used for the electrocatalytic reduction of $CO_2$. The effects of $[N_{1118}][TRIZ]$ concentrations and the water content on the electrocatalytic reduction of $CO_2$ with Ag as cathode were investigated separately through potentiostatic electrolysis. More importantly, the detailed mechanism of the electrochemical reduction of $CO_2$ in $[N_{1118}][TRIZ]$ ILs was investigated through density functional theory (DFT). This work aims to provide a new strategy to design novel ILs for high efficiency electrochemical reduction of $CO_2$ to CO.

## 2. Experimental section

*2.1. Materials and ILs synthesis*

1,2,4-Triazolylsodium (TrizNa) and Trimethyloctylammonium chloride ($[N_{1118}]Cl$) were purchased from Shanghai Aladdin Bio-Chem Technology Co., Ltd. Acetonitrile (AcN) was purchased from Fuchen Chemical Reagent Store, Xing'an Market, Anci District, Langfang City. Ag (purity>99.99%) electrode with diameter 1.0 cm was purchased from Tianjin Aida Hensheng Technology Development Co. Ltd. Gaseous $CO_2$ was purchased from Beijing beiwen Gas Factory. Gaseous $N_2$, $H_2$, and CO were obtained from Beijing Chengxin Shunxing Special Gas Technology Co., Ltd.

$[N_{1118}][TRIZ]$ was synthesized by dissolving equimolar $[N_{1118}]Cl$ and TrizNa in ethanol and allowed to stir for 12 h under room temperature. The white solid was removed through a Buchner funnel, and the filtrate was evaporated to remove ethanol at 65 °C. The obtained yellow viscous liquid was placed in a vacuum oven and dried at 75 °C for 24 h. White solid precipitated out when $CO_2$ gas was passed into the light yellow ILs dissolved by AcN. The white solid was removed and AcN was removed by rotary evaporation of the filtrate at 75 °C. The obtained light yellow viscous liquid was dried in a vacuum oven at 75 °C for 48 h. The $[N_{1118}][TRIZ]$ was analyzed using $^1$H-NMR and $^{13}$C-NMR using Bruker ARX-600.

*2.2. Electrochemical Measurements*

The electrochemical data were recorded employing in CHI-660E (Shanghai CH instruments Co., China) with Ag plate as working electrode, platinum gauze as counter electrode, and 0.01 mol $L^{-1}$ $Ag/Ag^+$ (formed by dissolving $AgClO_4$ in 0.1 M TBAP-AcN) as reference electrode. Cyclic voltammograms (CV) experiments were recorded in stand three-electrode configuration with a sweep rate of 50 mV/s from -1.3 V to -3.0 V (vs. $Ag/Ag^+$).

*2.3. Product analysis of the electrochemical reactions*

The electrochemical reactions were carried out at 293.15 K and atmospheric pressure in a H-type cell with Nafion 117 membrane separating the cathode and anode compartments. Prior to electrocatalysis test, the preparation consisted of polishing Ag electrode with sandpaper and rinsing with acetonitrile for more than 5 minutes. ILs was used as catholyte, and 0.1 M $H_2SO_4$ aqueous solution was used as anolyte. The catholyte was saturated by purging with $CO_2$ for at least 40 min before the $CO_2$ catalysis experiment. In order to mix better, the cathode solution was slight stirred while $CO_2$ gas was maintained at a speed of 60 ml $min^{-1}$. The gas products were collected using gas collecting bags and then analyzed by Agilent 7890B gas chromatography (GC) equipped with a thermal conductivity detector (TCD) and a flame ionization detector (FID) detector employing in $N_2$ as carrier gas. The definition of Faradaic efficiency (FE) was given in Eq. (1):

$$FE = nzF/Q * 100\% \qquad (1)$$

Where n is moles of electrons required to reduce the $CO_2$ to the product, F stands for the Faraday constant (96485 C/mol), z refers to the number of electrons involved, and Q represents the total charge moved through the reaction process.

*2.4. DFT calculations*

All DFT calculations were performed *via* the Gaussian 09 program package [18]. The Becke's three-parameter with Lee, Yang, and Parr's (B3LYP) [19] hybrid functional was applied to perform geometry optimization between ILs and $CO_2$ with the 6-311+g(d,p) basis set [20]. Absence of negative frequencies confirmed that the optimized molecular structures are the minima.

# 3. Results and Discussion

*3.1. Electrochemical characterization and product analysis*

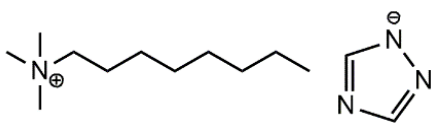

**Fig. 1.** Structure of the [$N_{1118}$][TRIZ]

With superior performance of ILs electrolyte in mind, ILs composed of special functionalized anion are candidates for research owing the unique structure, as show in **Fig. 1**. The result of NMR of [$N_{1118}$][TRIZ] showed that the synthesized ILs was consistent with the structure (Supporting Information, Fig. S1).

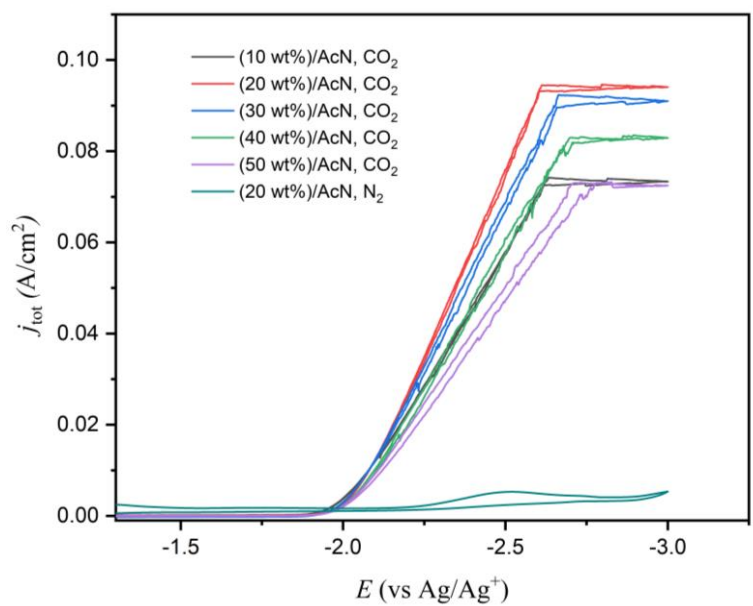

**Fig. 2.** CV curves with Ag electrode in the $N_2$-saturated and $CO_2$-saturated $[N_{1118}][TRIZ]$/AcN electrolyte.

A comparison of the cyclic voltammograms (CV) experiments were conducted to investigate the activity of $[N_{1118}][TRIZ]$ (**Fig. 2**). The high viscosity of ILs is not conducive to the reduction of $CO_2$. AcN, as aprotic organic solvents with very low viscosity, was often used in the electrochemical reaction process [1, 7]. Therefore, ILs/AcN mixtures were used in the study as electrolyte. When the $[N_{1118}][TRIZ]$/AcN were saturated with $N_2$ and $CO_2$ respectively, features changed in the CV responses occurred, as shown in **Fig. 2**. The significant measurements difference in the $N_2$-saturated and $CO_2$-saturated environment concerned the onset potential and the current density, indicating that the reduction of $CO_2$ occurs on the Ag electrode. Meanwhile, the content ILs $[N_{1118}][TRIZ]$ effected onset potential and the current density. It is worth to mention that the potential causing a current density of 0.6 mA cm$^{-2}$ was chosen as the onset potential of $CO_2$ reduction following the study of Lu et al. [21]. Therefore, the same onset potentials (-1.93 V (vs. Ag/Ag$^+$)) were found in different content of $[N_{1118}][TRIZ]$ saturated with $CO_2$ (**Fig. 2**), while the cathodic current density increased exponentially with $[N_{1118}][TRIZ]$ content increased from 10 wt% to 50 wt% at potentials < -1.93 V(vs. Ag/Ag$^+$). The current density tended to increase first and then decreased, and the highest current density for $CO_2$ reduction was obtained at the $[N_{1118}][TRIZ]$ content of 20 wt%. The main reason was that the ILs acted as a supporting electrolyte, resulting in different conductivity.

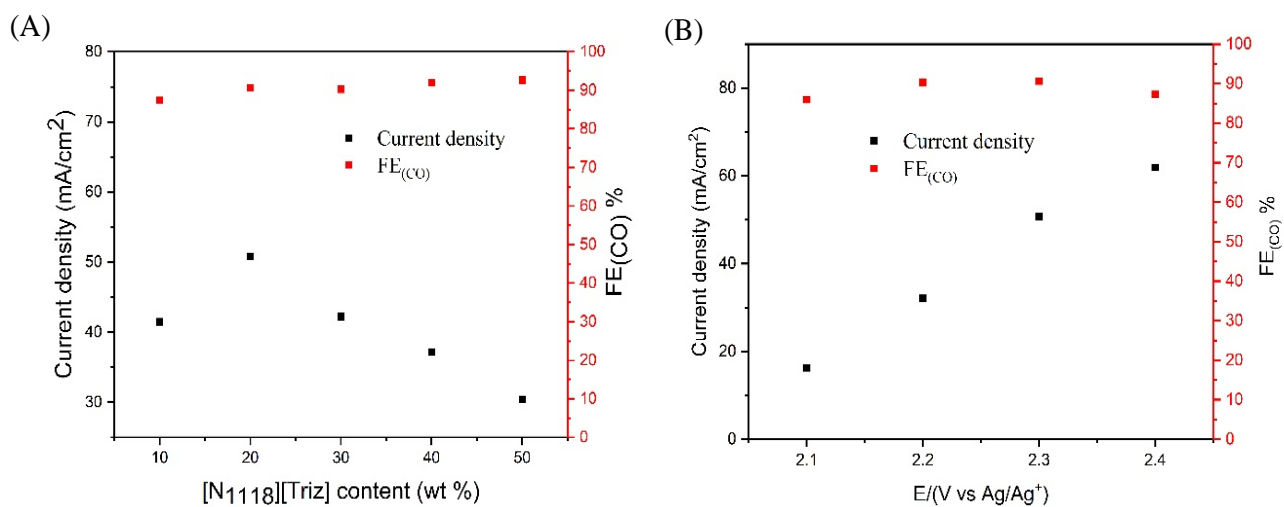

**Fig. 3.** (A) The effect of [N$_{1118}$][TRIZ] content on the current density and Faradaic efficiency with Ag electrode at -2.3 V (vs. Ag/Ag$^+$); (B) The effect of applied potential on Ag cathode in [N$_{1118}$][TRIZ](20 wt%)/AcN electrolyte at -2.3 V (vs. Ag/Ag$^+$).

The effect of [N$_{1118}$][TRIZ]-containing in binary electrolytes [N$_{1118}$][TRIZ]/AcN on the electrocatalytic activity of CO$_2$ reduction and the product selectivity were explored with Ag electrodes at -2.3 V(vs. Ag/Ag$^+$) for 2 h, as illustrated in **Fig. 3(A)**. As demonstrated in **Fig. 3(A)**, over 10 wt% [N$_{1118}$][TRIZ] content, the CO efficiency never fell below 90%. Clearly, the predominant CO formation prevented the formation of by-products H$_2$ in the [N$_{1118}$][TRIZ] electrolyte. When the [N$_{1118}$][TRIZ] contents were more than 20 wt%, the current density decreased. The reason was that electrostatic attraction between the cation and anion of ILs, causing the movement of ions to be blocked and the decrease of charge transfer rate. Remarkably, the current density was up to 50.8 mAcm$^{-2}$ with a Faradaic efficiency of 90.6% at -2.3 V (vs. Ag/Ag$^+$) in [N$_{1118}$][TRIZ](20 wt%)/AcN electrolyte. As mentioned previously, the current density obtained by using the [N$_{1118}$][TRIZ](20 wt%)/AcN electrolyte was much higher than those reported in the ILs electrolyte (**Fig. 3(A)**; Supporting Information, Table S1). These results showed that the binary [N$_{1118}$][TRIZ]/AcN mixtures yielded high total current densities and high CO selectivities. The corresponding potential dependent in the [N$_{1118}$][TRIZ](20 wt%)/AcN was conducted using Ag as electrode, as shown in **Fig. 3(B)**. The data showed that the Faraday efficiency exceeded 90% when the applied potentials were -2.2 V (vs. Ag/Ag$^+$) and -2.3 V (vs. Ag/Ag$^+$) respectively.

Water as important proton source provided the protons needed for $CO_2$ electrocatalytic reduction. A small amount of $H_2O$ in the process of electrocatalytic $CO_2$ reduction is beneficial to the reduction of $CO_2$ [7, 22]. Hence, the effect of water on electrocatalytic $CO_2$ reduction was investigated (**Fig. 4**). Herein, the water content was calculated based on AcN. **Fig. 4(A)** showed the CVs for the reduction of $CO_2$ on [$N_{1118}$][TRIZ](20 wt%)/AcN-$H_2O$ electrolyte. As the addition of water, it could be observed that the onset potential for the $CO_2$ reduction gradually moved to more positive potentials relative to the $CO_2$-saturated [$N_{1118}$][TRIZ]/AcN electrolyte (**Fig. 3(A)**). The onset potential of $CO_2$ reduction increased from -1.93 V to -1.75 V vs. Ag/Ag$^+$, as the water content increased from 0 wt% to 7.5 wt%. Lowering the initial reduction energy barrier upon water addition may be explained by increased the availability of proton near Ag electrode surface, which was beneficial to the $CO_2$ reduction [11]. **Fig. 4(B)** exhibited that the current density increased sharply with up to 7.5 wt% $H_2O$ content. However, the selectivity of CO decreased obviously (**Fig. 4(B)**), and it increased significantly for $H_2$, indicating that addition of too much water was to the disadvantage of $CO_2$ reduction. At the same time, Faraday efficiency of 86.2% with the 5 wt% $H_2O$ was obtained, and the current density of 61.7 mA cm$^{-2}$ was slightly lower than the current density at the water content of 7.5 wt%. The above analysis results showed that the high water content started to dominate the product distribution. This is consistent with literature [11, 14]. Considering better current density and Faraday efficiency with a water content of 5 wt%, we further explored the effect of applied potential in the [$N_{1118}$][TRIZ](20 wt%)/AcN-$H_2O$(5 wt%) mixtures.

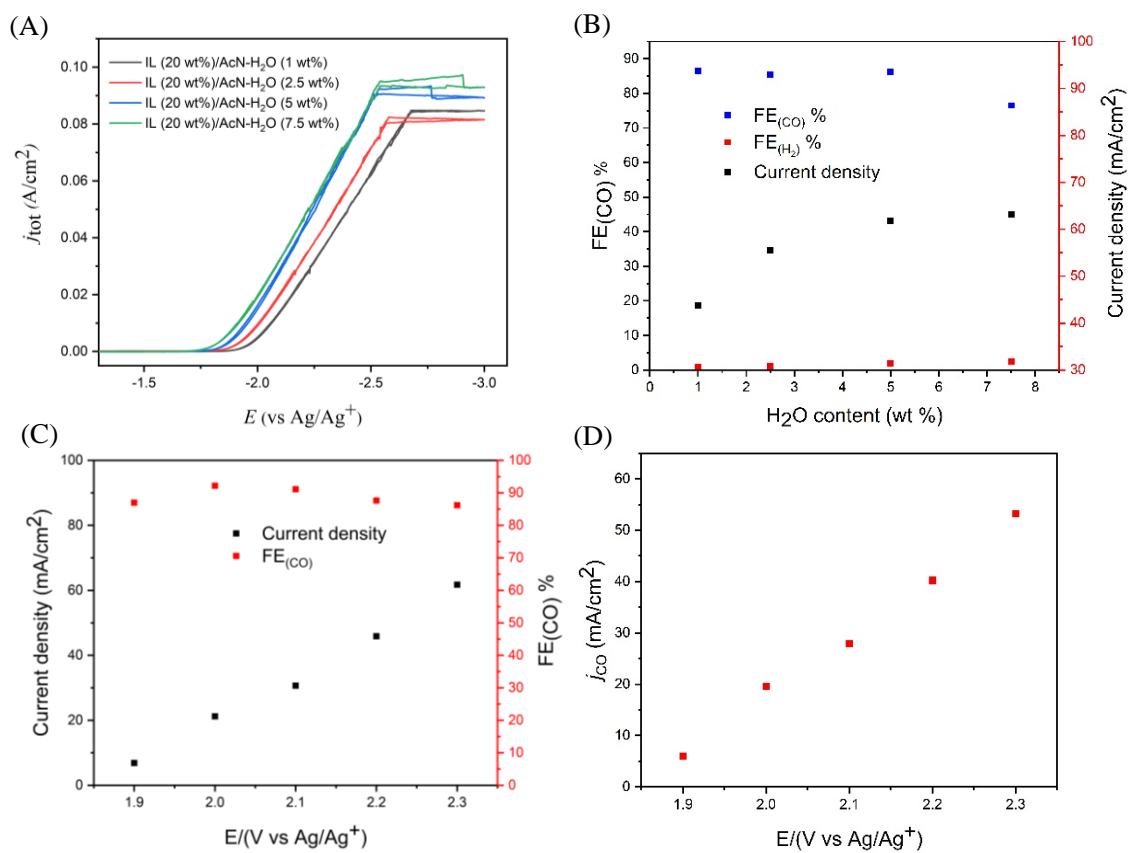

**Fig. 4.** The effect of adding $H_2O$ to $[N_{1118}][TRIZ]$(20 wt%)/AcN on $CO_2$ reduction on Ag cathode. (A) CVs measurements; (B) $H_2O$ amount; (C) The effect of applied potential on Ag cathode in $[N_{1118}][TRIZ]$(20 wt%)/AcN-$H_2O$(5 wt%) electrolyte at -2.3 V (vs. Ag/Ag+); (D) Partial current density of CO in different applied potential.

The effect of applied potential in the $[N_{1118}][TRIZ]$(20 wt%)/AcN-$H_2O$(5 wt%) mixtures illustrated in **Fig. 4(C)**. Obviously, the total current density and CO partial current density (**Fig. 4(D)**) increased with the increased of applied potential. At medium applied potentials ranged from -2.0 V to -2.1 V (vs. Ag/Ag$^+$), the CO efficiency higher than 91%. At -2.0 V (vs. Ag/Ag$^+$), the Faraday efficiency of CO achieved a maximum value of 92.2% with a current density of 21.2 mAcm$^{-2}$. The above results indicated that the ILs helped to provide a low energy with high current density and Faraday efficiency for $CO_2$ electroreduction.

*3.2. DFT Calculations*

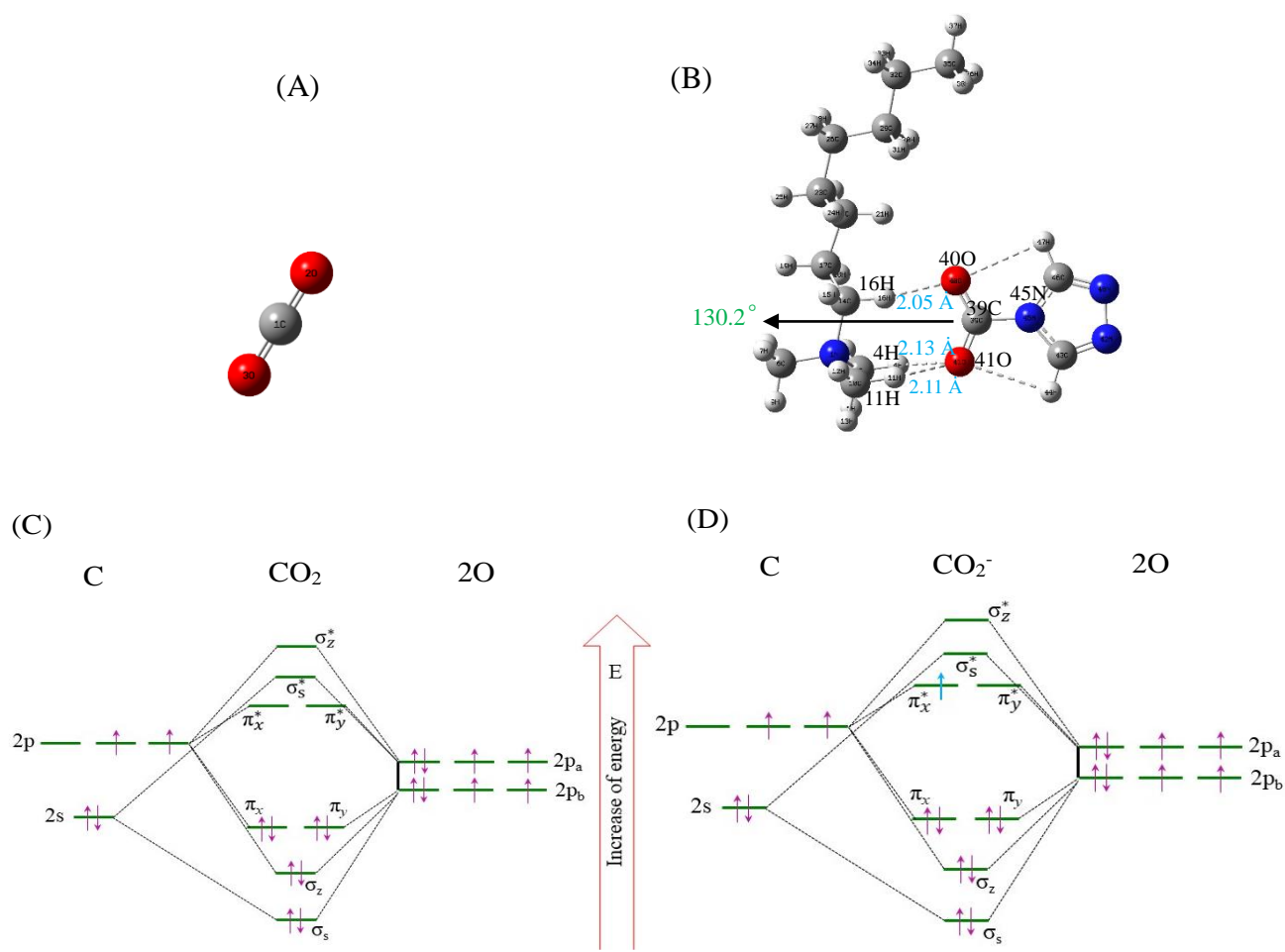

**Fig. 5.** The lowest energy configurations: (A) $CO_2$; (B) $[N_{1118}][TRIZ]$-$CO_2$. N atom (blue), C atom (gray), H atom (white). (C) Energy level diagram of $CO_2$ molecular orbital; (D) Energy level diagram of after $CO_2$ activation.

**Table 1**

Van Der Waals radius rw (nm) of atoms [23]

| C | H | Br | Cl | N | O | P | S |
|---|---|----|----|---|---|---|---|
| 1.72 | 1.2 | 1.85 | 1.75 | 1.55 | 1.5 | 1.8 | 1.8 |

**Table 2**

DFT calculation results of $CO_2$ and $[N_{1118}][TRIZ]$-$CO_2$

|  | C=O bend length (Å) | Bond angle (°) |
|---|---|---|
| $CO_2$ (**Fig. 5(A)**) | 1.16 (2O---1C)<br>1.16 (3O---1C) | 180 |
| $[N_{1118}][TRIZ]$-$CO_2$ (**Fig. 5(B)**) | 1.24 (40O---39C)<br>1.25 (41O---39C) | 130.2 |

The quantum chemical calculations were carried out to explore the underlying mechanism of $[N_{1118}][TRIZ]$ activating $CO_2$. As was evident in **Fig. 5(B)**, the most stable configuration between $[N_{1118}][TRIZ]$ and $CO_2$ was obtained by structure optimization. The foremost feature of hydrogen bond was that the proton–acceptor distance was shorter than the corresponding the sum of their van der Waals radii [23] (**Table 1**). The calculation result in **Fig. 5(B)** showed that the interatomic distances of 40O⋯16H (2.05 Å), 41O⋯4H (2.13 Å), and 41O⋯11H (2.11 Å) represented strong hydrogen bond interactions, including that the cation formed hydrogen bonds with $CO_2$. Furthermore, the hydrogen bonding strength of 40O⋯16H (2.05 Å) compared with those 41O⋯4H (2.13 Å) and 41O⋯11H (2.11 Å) was much stronger (**Table 2**). Moreover, N atom at the 4-position (45N) of 1,2,4-triazole preferred to coordinate with C atom of $CO_2$ because of its strong electronegativity, resulting in a formation of [TRIZ-$CO_2$]$^-$ complex.

**Table 3**

NPA charge of $CO_2$ and NPA charge of $CO_2$ in [$N_{1118}$]$^+$[TRIZ-$CO_2$]$^-$ complex

| $CO_2$ | | [$N_{1118}$]$^+$[TRIZ-$CO_2$]$^-$ | |
|---|---|---|---|
| Atom | Net charge | Atom | Net charge |
| C | 0.99041 | C(39) | 0.46314 |
| O | -0.49520 | O(40) | -0.37486 |
| O | -0.49520 | O(41) | -0.39492 |
| Total charge | 0.00000 | Total charge | -0.30664 |

The Natural Bond Orbital (NBO) analysis was performed, aiming at getting more detailed information of $CO_2$ activation process (**Fig. 5(C)**, **Fig. 5(D)** and **Table 3**). Based on molecular orbital theory, the $CO_2$ molecular bonding orbital consisted of the double electrons occupied orbitals ($\sigma$ and $\pi$) and the electron unoccupied antibonding orbitals ($\sigma^*$ and $\pi^*$) (**Fig. 5(C)**), as such the $CO_2$ molecule was relatively stable. Particularly, according to Natural Population analysis (NPA) in the [$N_{1118}$]$^+$[TRIZ-$CO_2$]$^-$ complex, the net charge of $CO_2$ is -0.307 e compared to the isolated neutral $CO_2$ molecule (**Table 3**), which showed that $CO_2$ got electrons from [TRIZ]$^-$ anion. The additional electron of $CO_2$ was filled in the Lowest Unoccupied Molecular Orbital (LUMO), which was mainly composed of the anti-bonding orbit of $CO_2$. Hence, that is to say, the obtained electron of $CO_2$ from [TRIZ]$^-$ preferred to occupy the anti-bonding orbit ($\pi^*$) (**Fig. 5(D)**), thus reducing the electron density between C and O in $CO_2$. On the basis of molecular orbit theory, the higher $\pi^*$ occupancy or $\sigma^*$ occupancy, the higher energy of the system, thus weakening the chemical bond and reducing the bond energy [24].

Therefore, C=O bond energy could be weakened owing to the introduction of anti-bonding orbits electrons. This is also the reason why the C=O changed from 1.16 Å to 1.24 Å and 1.25 Å (**Table 2**). It could be seen from **Fig. 5(B)** that the linear spatial distribution of $CO_2$ molecule (C, sp hybridization) altered to the bent configuration (C, near $sp^2$ hybridization) with O-C-O bond angle of 130.2° after the action of $[N_{1118}][TRIZ]$. At the same time, the $CO_2$ molecule was lengthened asymmetrically from the original symmetric C=O, as shown in **Table 2**. The above analysis demonstrates that $[N_{1118}][TRIZ]$ is able to provide a low energy path for $CO_2$ electrocatalytic reduction.

*3.3. Catalytic mechanism*

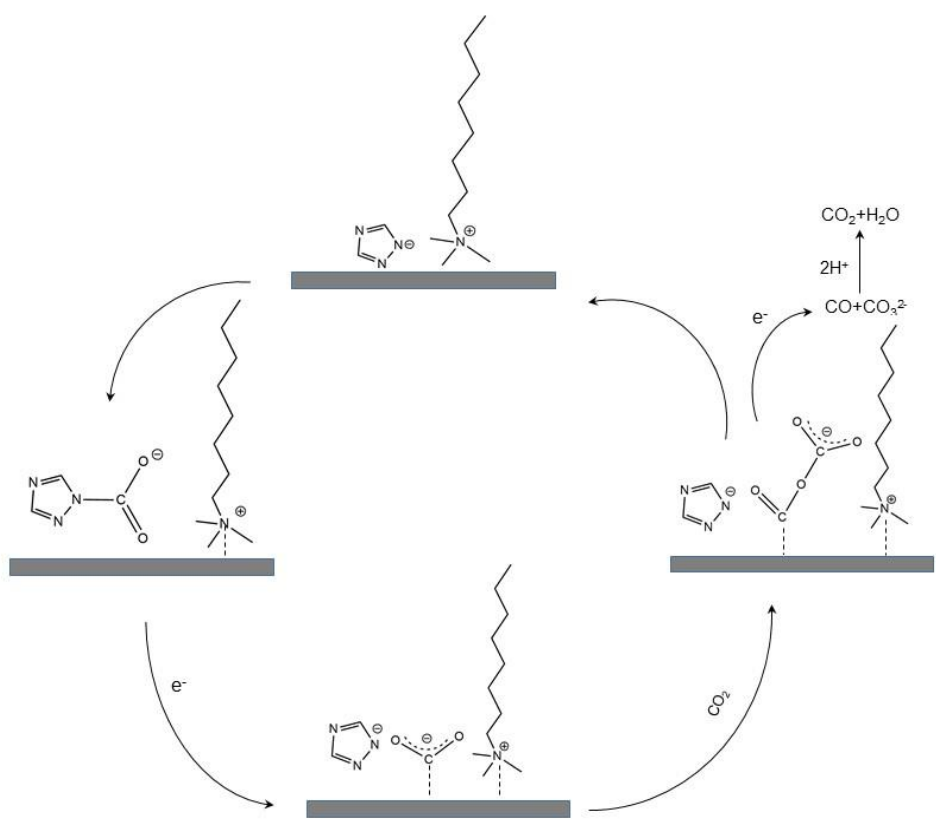

**Fig. 6.** Possible mechanism of electrochemical reduction of $CO_2$ to CO on Ag cathode in $[N_{1118}][TRIZ]$-containing catholytes.

The mechanism of electrocatalytic reduction of $CO_2$ to CO in imidazole based ILs electrolyte had been studied by many researches [6, 11]. Imidazole cations play a key role in this kind of electrolyte. However, unlike conventional imidazole-based ILs, the $[N_{1118}][TRIZ]$ mainly affects the reaction process by anion. Hence, a mechanism different from imidazolium-based ILs was proposed for the electrocatalytic $CO_2$ reduction to CO on Ag cathode in $[N_{1118}][TRIZ]$-containing catholytes (**Fig. 6**) based on the experimental studies in the open literature [25-27]. From Section 3.2, the $CO_2$ molecule receives extra electron from the $[TRIZ]^-$, leading to more higher $\pi^*$ occupancy, which caused $CO_2$ to change from linear to bend. Formation of $[N_{1118}]^+[TRIZ-CO_2]^-$ complex could lower the reaction energy barrier of $CO_2$ reduction. Subsequently, the $[N_{1118}]^+[TRIZ-CO_2]^-$ complex was adsorbed on the surface of the Ag electrode and formed $[N_{1118}]^+[TRIZ-CO_2]^-$ layer. The surface of Ag electrode was enriched with electrons under the effect of applied potential, and an electron from Ag surface tended to transfer to the anti-bonding orbital ($\pi^*$) of the activated $CO_2$ by $[N_{1118}][TRIZ]$. As a result, the bent $CO_2$ got an electron from the donor to form an adsorbed $CO_2^{\bullet-}$ radical on the surface of Ag. After that, $CO_2^{\bullet-}$ radical combined with a $CO_2$ to generate adduct of $(CO_2)_2^{\bullet-}$. Finally, $(CO_2)_2^{\bullet-}$ formed CO by obtaining an electron. The product CO diffused into bulk solution through the gap between the adsorption ILs layers.

## 4. Conclusion

In this work, we presented a new ILs $[N_{1118}][TRIZ]$ for electrocatalytic $CO_2$ reduction on a commercial Ag electrode. The result demonstrated that the $[N_{1118}][TRIZ]$/AcN system exhibits excellent performance for $CO_2$ electroreduction to CO with current density of up to 50.8 $mAcm^{-2}$ and a CO Faradaic efficiency of 90.6%. The $[N_{1118}][TRIZ]$ (20 wt%)/AcN-$H_2O$ (5 wt%) mixture electrolyte showed that the Faradaic efficiency of 92.2% with a current density of 21.2 $mAcm^{-2}$ could be obtained. These excellent effects came from the activation of $CO_2$ by $[N_{1118}][TRIZ]$. According to NBO analysis, the electrons obtained from $[N_{1118}][TRIZ]$ would be filled in the anti-bonding orbit ($\pi^*$) of the $CO_2$, thus reducing the reduction barrier required for electrocatalytic $CO_2$ reduction. From the achievements in this contribution, we believe that it is possible to conduct a commercial process for electrochemical $CO_2$ reduction with the help of the new ILs-based electrolyte in combination with non-precious metal catalysts.

# Acknowledgements

This work is financially supported the program of Beijing Municipal Natural Science Foundation (2182072, 2182071).

# Appendix A. Supplementary data

Supplementary material related to this article can be found in the online version.